# Options for Managing Spent TRi-structural ISOtropic Nuclear Fuel

Harish Gadey[*], Stuart Arm[*], Justin Clarity[*], Pavlo Ivanusa[*], Stephen Davidson[*], Drew Fairchild[**], Robert Pierce[**], Ernest Hardin[***]

[*]Pacific Northwest National Laboratory, Richland, WA, USA
[**]Savannah River National Laboratory, Aiken, SC, USA
[***]Longenecker & Associates, Inc., Oak Ridge, TN

**ABSTRACT**

During the past few decades, there has been a renewed interest in advanced reactor concepts, especially the high-temperature gas-cooled and the molten salt-cooled technologies for energy production as well as hydrogen or process heat generation. Tri-structural isotropic (TRISO) fuel is intended to be used in some of these emerging reactor technologies. The fuel bearing TRISO particles are encapsulated in graphite, which acts in the capacity of a moderator. TRISO based fuels can be fabricated in the form of spherical pebbles or cylindrical compacts, the latter of which are then loaded into prismatic graphite blocks.

A few TRISO fuel-based reactor concepts were initially analyzed to estimate the volumetric waste generated to inform waste management related activities. Volumetric discharges ranged between 4.4 to 16 times that of conventional pressurized water reactors (PWRs). However, close to an order of magnitude reduction in the waste volume is estimated for some reactor concepts (using a 25% packing fraction) if the TRISO particles can be successfully separated from the graphite moderator so the latter may be potentially disposed of as low-level radioactive waste (LLW).[1]

This was followed by using publicly available data from X-Energy's Xe-100 reactor (such as 173 pebbles discharged per day) to explore the spent nuclear fuel (SNF) characteristics and management methods within the currently conceived waste management system. The Standardized Computer Analyses Licensing Evaluation (SCALE) system of codes was used to estimate the decay heat as a function of time based on burnup, enrichment, and specific power. The fission product contamination in the graphite moderator was estimated based on data available from prior research and its implications on disposal as LLW. From a separation standpoint, the SNF containing the graphite and the TRISO particles is estimated to be stored for about 100 years to allow for the fission products to decay sufficiently to reasonably attempt separation. It is estimated that after about 100-years post discharge, a maximum of 5 TRISO particles out of a total of 19,000 TRISO particles inside a single Xe-100 reactor pebble can be tolerated in the graphite. In other words, 100 years post discharge, if greater than 5 TRISO particles are left in the pebble post-separation, the graphite moderator would potentially not meet the LLW discharge criteria. This does not consider in-service failure of the TRISO particle resulting in contamination during the operation of the reactor. Achieving this high level of TRISO particle separation from the graphite at scale is likely to prove to be a significant technical challenge.

---

[1] Legal issues, including those concerning the classification and disposal of SNF or LLW, are beyond the scope of this technical paper.

Packaging and transportation analysis was also performed on separated TRISO particles as well as intact TRISO SNF. For this analysis, it was assumed that the TRISO particles were encapsulated in a medium such as borosilicate glass or metal. TRISO loading fractions were varied between 20 and 50 percent to account for uncertainty. The radiation dose analysis from the package was found to be compliant for normal conditions of transport and hypothetical accident conditions (As per Title 10 of the Code of Federal Regulations (CFR) Part 71) within a timeframe comparable to light water reactor (LWR) SNF. From a package inventory perspective, for intact TRISO SNF, the various TRISO reactor designs considered require between 6.5 and 13.5 times the number of dual-purpose canisters (DPCs) equivalent compared to a typical LWR fuel cycle. However, for separated TRISO SNF, a DPC usage reduction factor between 4.2 and 23.8 times was observed for various TRISO loadings (by volume) between 20 and 50 percent. In addition, a criticality analysis using several TRISO loading fractions in borosilicate glass and CERMETs is also presented in this work.

**INTRODUCTION**

Water cooled reactors have been deployed all over the world for the past seven decades and represent a well-established technology. High-temperature gas-cooled reactors and molten salt technologies have been gaining traction in the past decade. In addition to fulfilling energy needs these technologies might also support process heat generation for the manufacturing industry and hydrogen production among other alternate uses. TRISO fuel is anticipated to be used in some of these emerging reactor design concepts. TRISO fuel particles are generally encapsulated in graphite that acts in the capacity of a moderator. TRISO fuel may be fabricated in the form of pebbles or cylindrical compacts, the latter of which may be loaded in prismatic graphite blocks. This paper focuses on three advanced reactor concepts: X-Energy's Xe-100, Kairos Power's Fluoride Salt-Cooled High-Temperature Reactor (KP-FHR), and Framatome's Steam Cycle High-Temperature Gas-Cooled Reactor (SC-HTGR).

One of the potential advantages of using TRISO fuel is increased burnup. The main in-service life-limiting component for traditional LWR reactors is the fuel cladding. This further motivated the development of coated particle fuel concepts (including TRISO) to increase fission fragment retention efficiency. TRISO fuel is typically made of approximately 1 millimeter (mm) diameter particles that may be manufactured into compacts of cylindrical shape (approximately 12 mm diameter and 25 mm length) or spheres ("pebbles") approximately 60 mm diameter. Compacts are fabricated by mixing graphite powder with a phenolic resin, then adding the TRISO particles and isostatically pressing the mixture. The compacts are sometimes subjected to a sintering process to increase fuel density. The spherical pebbles, on the other hand, are coated with graphite (approximately 5 mm thickness) and would be used to fuel a pebble-bed reactor (PBR). An individual TRISO fuel particle is shown in Figure 1.

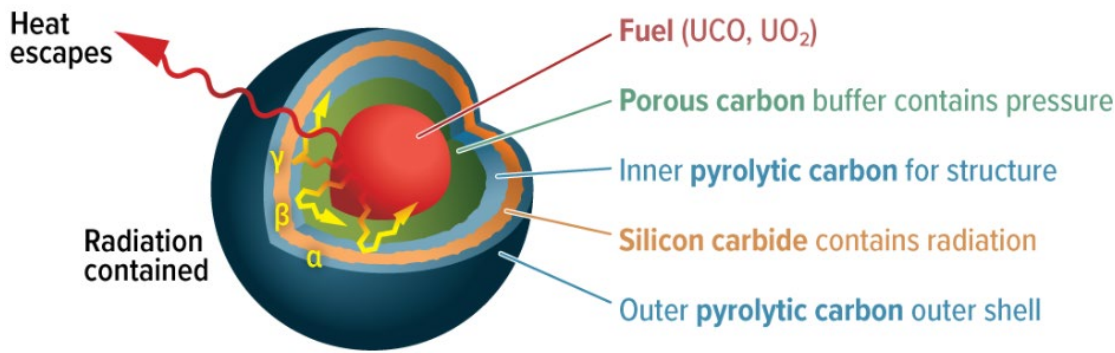

Figure 1. Layers of a Tri-structural isotropic (TRISO) fuel particle [1].

**VOLUMETRIC WASTE GENERATION AND DECAY HEAT ESTIMATES**

Estimating the waste volume being generated is an important factor in understanding implications for SNF management. In this analysis, three reactors were considered, the Xe-100, KP-FHR, and the SC-HTGR. In addition, for comparison, data from traditional PWRs, and boiling water reactors (BWRs) is also presented (Table 1). This table captures publicly available data for each reactor's electrical power output, discharge volume per year, discharge mass per year, and pebble packing fraction (if applicable).

Table 1. Summarized Discharge Values for Various Reactor.

| Reactor | Xe-100 [2] | KP-FHR [3, 4, 5] | SC-HTGR [6] | PWR [7] | BWR [7] |
|---|---|---|---|---|---|
| Electrical Output (MWe) | 80 | 140 | 272 | 984 | 1002 |
| Discharge Volume ($m^3$/yr) | 7.00 | 3.37 | 34.82 | 9.1 | 10.6 |
| Discharge Mass (MT/yr) | 12.70 | 5.87 | 54.29 | 31.7 | 40.0 |
| Discharge Volume/Electric Output [($m^3$/yr)/MWe] | 8.75E-02 | 2.41E-02 | 1.28E-01 | 9.25E-03 | 1.06E-02 |
| Discharge Mass/Electric Output [(MT/yr)/MWe] | 1.59E-01 | 4.19E-02 | 2.00E-01 | 3.22E-02 | 3.99E-02 |
| Pebble Packing Fraction (%) | 60 | 60 | – | – | – |

Discharge decay heat is another important factor that helps characterize the SNF from advanced reactors. The decay heat characteristics were modeled using the Oak Ridge Isotope Generator (ORIGEN) tool for the Xe-100 reactor concept. Figure 2 presents the volumetric decay heat for Xe-100 (PBR) as well as for a traditional LWR. Volumetric decay heat ratio values for Xe-100 and the LWR concept were also provided as a function of time. For the Xe-100 reactor concept, a burnup of 165 Megawatt-day per metric ton Uranium (MWd/MTU), fuel enrichment of 15.5%, and a specific power of 129.6 Megawatt per metric ton Uranium (MW/MTU) was used. The analysis indicated that although some advanced reactor concepts are anticipated to generate a large amount of SNF from a volumetric standpoint, the volumetric decay heat is a fraction when compared to LWR SNF.

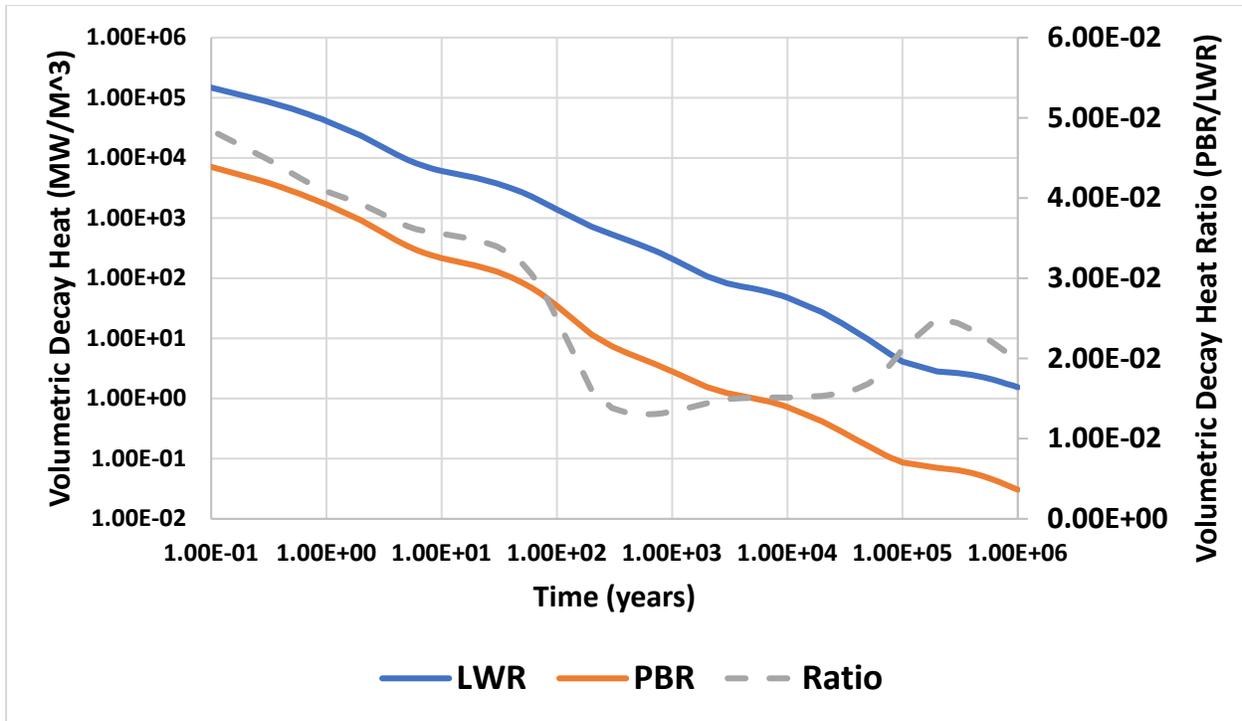

**Figure 2. Volumetric decay heat as a function of time for PBR and LWR reactor.**

**TRISO PARTICLE SEPARATION FOR LLW CLASSIFICATION**

An attempt was made to estimate the TRISO particle separation requirements to classify, from a technical standpoint, the bulk graphite as low-level waste. For these estimations, the pebble dimensions provided by X-Energy were referenced; wherein the fuel particle diameter was 0.0425 centimeter (cm), single particle diameter was 0.0855 cm, and the pebble diameter was 6 cm. In addition, as per data from X-Energy, approximately 19,000 particles per pebble and 220,000 pebbles per core were used to estimate the total composition of the core [2, 8]. Based on these values, the total volume of fuel particles in the reactor was calculated to be 0.168 cubic meters ($m^3$), the total volume of graphite in the core was 23.51 $m^3$, and the volume fraction of graphite in a core was estimated to be 0.945. This yields a total of about 47,027 kg of graphite in the core. The density of the fuel particle was assumed to be 10.5 grams per cubic centimeter ($g/cm^3$) and the density of the pyrolytic carbon (and graphite) was estimated to be 2 $g/cm^3$. Combining the volume and densities provided a fuel particle mass fraction in the core to be 0.42.

10 CFR Part 61 lists the different classes of LLW as well as the concentration limits for each class. For alpha-emitting transuranic radionuclides, the concentration limit is 100 nanocurie per gram (nCi/g). Using this limit, TRISO particle decontamination factors were calculated. A TRISO particle decontamination factor of over 4,000 was calculated after aging the SNF for 100 years as shown in Table 2. To aid with visualization, these decontamination factors were extended to calculate tolerable numbers of SNF particles per core and SNF particles per pebble. In this context, the term "tolerate" refers to the maximum number of SNF particles that may be present (on a per core or pebble basis) after separation to be classified as LLW. After aging the SNF for 10 years, about 101,000 SNF particles can be tolerated from an entire core of 220,000 pebbles, or less than 1 particle per pebble on average. After aging the SNF for 100 years about 900,000 SNF particles can be tolerated or approximately a little over four SNF particles out of 19,000 per pebble.

Table 2. Number of tolerable particles per core and per pebble for LLW classification using the Alpha-Emitting Transuranic limit of 100 nCi/g.

|  | Parameter Value | | | |
| --- | --- | --- | --- | --- |
| Parameter | 0 years | 10 years | 100 years | 200 years |
| Limit [nCi/g] | 100 | 100 | 100 | 100 |
| Concentration in discharged intact SNF [nCi/g] | 3,169,214,519 | 4,138,226 | 452,859 | 269,650 |
| Required particle decontamination factor | 31,692,145 | 41,382 | 4,529 | 2,697 |
| # Particles tolerable per core | 131.89 | 101,009.5 | 923,024.6 | 1,550,158 |
| # Particles tolerable per pebble | 0.0006 | 0.459 | 4.195 | 7.046 |

From a technical standpoint, removing all but 5 particles from a deconsolidated pebble is likely very challenging and requires a reliable and robust quantifiable industrial process which can demonstrate that the limit can be satisfied.

**PACKAGING AND TRANSPORTATION ANALYSIS**

This section discusses aspects of radiation dose, criticality analysis, as well as the packaging inventory requirements for both intact SNF (moderator and TRISO particles) and separated TRISO particles.

**Criticality Analysis**

The packaging systems that will be used to transport TRISO SNF are currently not finalized. Intact TRISO SNF was evaluated to be less limiting compared to LWR SNF from a structural, thermal, and radiation dose perspective. Therefore, demonstrating subcriticality of the packaging system became the primary design consideration. Because of the lack of a regulatory basis for burnup credit for TRISO SNF, a fresh fuel criticality safety analysis was used to size the canisters. The Xe-100 and KP-FHR are pebble fueled reactors and are more amenable to a sub-canister concept (packing fraction 0.61), while the SC-HTGR fuel is in the form of prismatic blocks and is better accommodated by a single canister with a hexagonal basket. SCALE 6.2.4 version of KENO-VI was used to perform these calculations. For the Xe-100 SNF, the largest system that could remain subcritical under the optimal water ingress scenario is the General-37[2] loaded with seven DOE 24 × 15 standard canisters. For the KP-FHR SNF, the canister design suggested by Blandford and Peterson could be demonstrated to be adequately subcritical [3]. The KP-FHR canister design is arranged as two layers of 19 sub-canister arrays, with each sub-canister having an inner diameter of 30.5 cm (12 inches). The largest package that could be used to transport the SC-HTGR SNF was determined to be a General-37 with 19 radial locations for SNF blocks. Each radial location can accommodate 5 axial blocks for a total canister capacity of 95 SNF blocks.

For separated TRISO particles, infinite medium criticality calculations were performed to demonstrate that the immobilized TRISO particles could be transportable from a criticality perspective. This approach does not consider neutron absorption in basket materials and neutron leakage. The criticality calculations considered variations in the TRISO loading in the waste form, the porosity of the waste form, and the immobilization material. The TRISO volume fraction was varied between 20 and 50 percent to understand the impact of increased loadings. Glasses are known to fracture when they are cooled, leading to the formation of void spaces within the glass matrix. Under transportation accident conditions, the void space could be filled by water. There is limited information about the volume of the void space that may be available in case of other immobilization materials like the CERMET. To understand each material's tolerance, void formation calculations were run modeling a range of materials consisting of TRISO particles as a mixture of water and immobilization medium in ratios varying from 0–90 percent water. Materials considered in this evaluation include borosilicate glass [9], CERMET [10], and CERMET with 2 weight percent natural gadolinium added to the material. The criticality calculations were performed with enough neutron histories to produce an uncertainty in $k_{eff}$ of <0.00030. The results of the criticality calculations in Figure 3 show that the there is significant criticality margin for the borosilicate glass as well as the poisoned CERMET cases regardless of loading or porosity level (<90 percent), while the non-poisoned CERMET showed a significant porosity dependence on criticality suppression.

---

[2] The General-37 model is intended to be a generic nonproprietary canister and cask system for storing used nuclear fuel. It is based on a modified version of the MPC-37 canister. The square baskets can hold 37 typical 17 × 17 PWR fuel assemblies. The canister is made of stainless steel. The basket which holds the fuel assembly is made of borated aluminum.

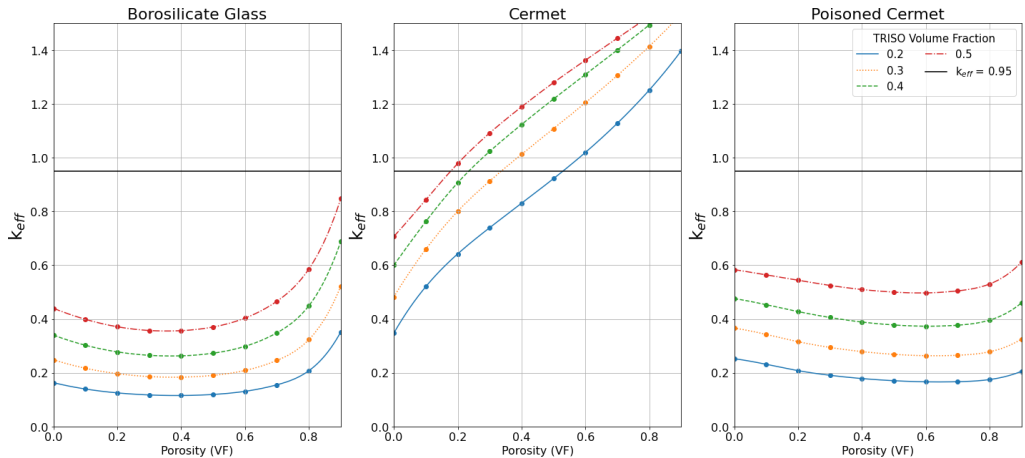

**Figure 3. $k_{eff}$ calculations for immobilized (separated) TRISO SNF particles as a function of medium porosity.**

**Package Dose Analysis**

Package dose is one of the important factors to consider for SNF storage and transportation. SCALE 6.2.4 version of MAVRIC was used to perform these calculations. The normal conditions of transportation (NCT) 2 m dose results[3] in Figure 4 show that a separated TRISO transportation package containing Xe-100 SNF would satisfy the 10 milli Roentgen equivalent man per hour (mrem/h) limit with approximately 3 years of cooling time for a 20 percent loading and approximately 4.5 years of cooling time for a 50 percent loading. The separated KP-FHR SNF would satisfy the 2 m dose limit with approximately 4.5 years of cooling time for a 20 percent loading and with approximately 7.5 years of cooling time for a 50 percent loading.

---

[3] A general-37 package using approximately 11.5 cm carbon steel, 8.5 cm lead and 16 cm of neutron shielding resin was used.

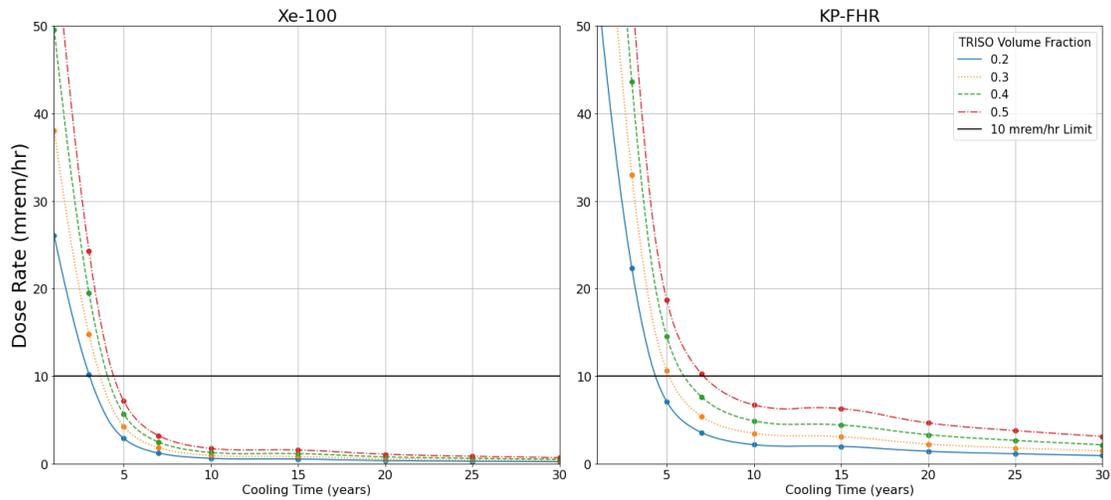

**Figure 4. Dose Rate at 2 m from a transportation package under NCT containing separated Xe-100 SNF (left) and KP-FHR SNF (right) as a function of cooling time.**

The hypothetical accident condition (HAC) 1 m dose results (Figure 5) show that the separated Xe-100 SNF would satisfy the 1000 mrem/h limit within less than a year of cooling time regardless of particle loading. The KP-FHR separated SNF would require approximately 4 years of cooling time to satisfy the HAC limit with a 50 percent loading and would require under 1 year of cooling time for lower loadings.

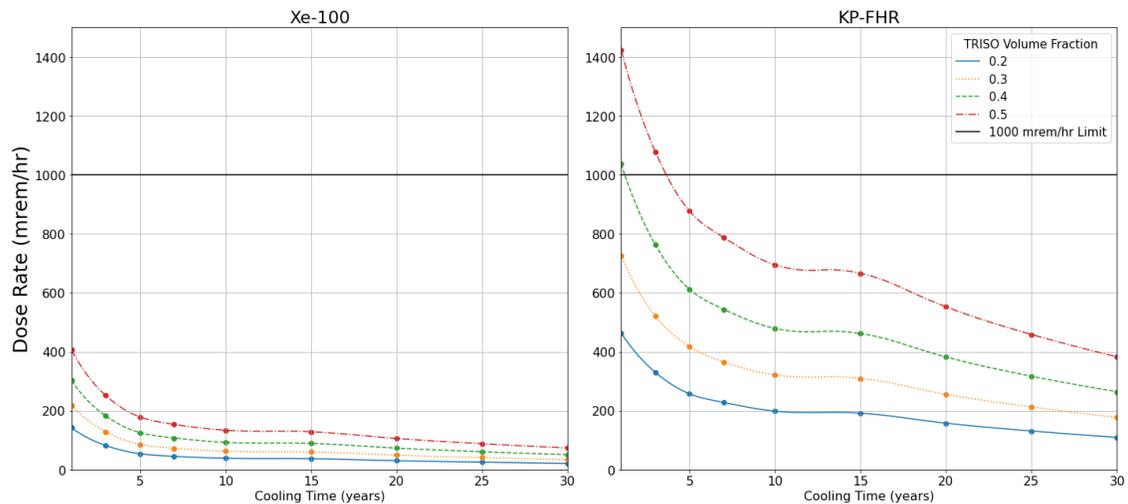

**Figure 5. Dose rate at 1 m from the transportation package surface during a HAC scenario for the separated Xe-100 SNF (left) and the KP-FHR SNF (right) as a function of cooling time.**

**Decay Heat Analysis**

For LWR SNF, the limiting phenomenon is the 400 Degree Celsius (ºC) peak clad temperature that is intended to prevent degradation of the zirconium-based cladding. Since TRISO fuel has a significantly different composition, this factor is not expected to be limiting from a thermal performance standpoint. Since the transportation package considered here is conceptual, for this work a representative 30 kilowatt (kW) decay heat limit was assumed for transportation analyses. The ORIGEN tool was used for decay heat estimations using various TRISO loading fractions. Figure 6 shows the total package decay heat as a function of cooling time for various TRISO loading fractions (separated SNF) for the Xe-100, and KP-FHR reactor SNF. For the Xe-100 reactor even at 50% loading fraction, the heat loads are below the 30-kW limit within 10 years. On the other hand, for the KP-FHR SNF, at the 10 years mark the heat loads are at/or below the limit for the 20 and 30% loading scenarios while for the 50% loading scenario, the 30-kW heat limit is expected to be met after about 35 years of cooling.

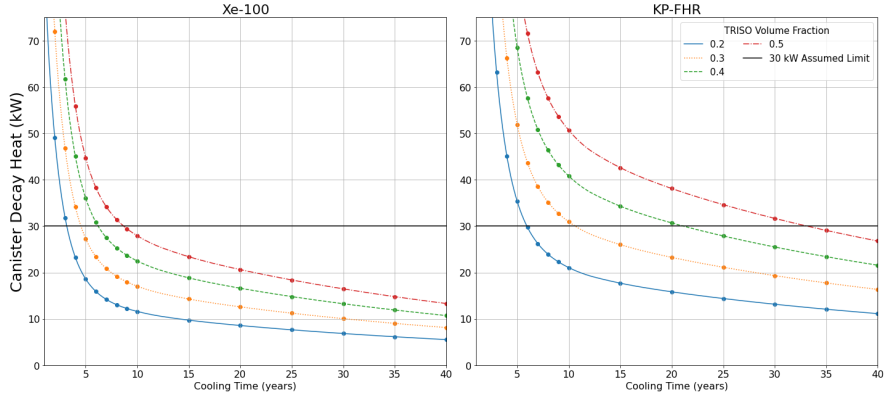

**Figure 6. Package decay heat for the separated Xe-100 (left) and separated KP-FHR (right) reactor SNF as a function of cooling time.**

**Package Requirement Analysis**

The analysis performed in this work shows that the number of DPC equivalent packages necessary to handle the volume of intact TRISO SNF in pebble form may be significantly higher than what is currently used for the LWR fuel cycle. However, this could be reduced substantially by separating the TRISO particles from the bulk graphite. The intact TRISO SNF for the Xe-100, SC-HTGR, and KP-FHR would require about 13.5, 10.9, and 6.5 times the number of DPC equivalent packages as the current LWR fuel cycle. In all cases the DPCs used were based on modern 37 PWR assembly canisters. The DPCs used for the comparison have an inner diameter of 1.88 m and an inner cavity height of 4.7 m. Table 3 shows the DPC-equivalent reduction factors that might be achieved for separated TRISO SNF particles. These reduction factors were calculated by dividing the number needed for the intact case by the number needed as a function of volumetric TRISO particle loading (for the separated cases). These results show that depending on the achievable TRISO particle loading, it may be possible to reduce the number of DPC equivalent packages needed to near LWR levels, or potentially even lower.

**Table 3. DPC Equivalent reduction factors for separated TRISO SNF particles.**

| TRi-structural ISOtropic Loading | Xe-100 | Steam Cycle High-Temperature Gas-Cooled Reactor | Kairos Power Fluoride Salt-Cooled High-Temperature Reactor |
|---|---|---|---|
| 20% | 7.2 | 5.8 | 4.2 |
| 30% | 10.5 | 8.5 | 6.2 |
| 40% | 13.9 | 11.2 | 8.2 |
| 50% | 23.8 | 19.2 | 14.1 |

**CONCLUSIONS**

This work explored several SNF characteristics of reactor concepts that are planning to use TRISO fuels in both compact (prismatic) and pebble form. Initially, the annual discharge characteristics of various reactors such as mass and volume were explored. The results from annual waste volume produced as a function of percentage annual nuclear energy generation was presented for the Xe-100, KP-FHR, and the SC-HTGR reactor concepts. The increase in waste volume was projected to be between 4.4 and 16 times compared to PWR reactors. Decay heat analysis using ORIGEN was performed to understand the volumetric decay heat characteristics of TRISO based and traditional LWR SNF. During the analysis time frame (up to 1 million years), it was observed that the ratio of TRISO volumetric decay heat fluctuated between 1.5% and 5% that of traditional LWR SNF. An attempt was made to understand the level of TRISO particle separation needed to classify the graphite moderator as LLW. This analysis yielded that at the 100-year mark, at-most 4 TRISO particles can be tolerated in a Xe-100 pebble for the graphite to be classified as LLW. Packaging and transportation analysis was also performed from a standpoint of criticality, package dose, decay heat, and approximate number of packages required to accommodate the SNF. Criticality studies were performed for both intact as well as separated TRISO SNF using a variety of immobilization media. Package dose analysis was performed for the Xe-100 and KP-FHR reactor concepts for both normal conditions of transportation as well as hypothetical accident conditions. Package decay heat analysis was performed assuming a 30-kW package heat limit for the Xe-100 and KP-FHR SNF assuming a variety of TRISO loading fractions between 20 and 50 percent. From a number of packages requirement standpoint, the intact TRISO SNF for the Xe-100, SC-HTGR, and KP-FHR would require about 13.5, 10.9, and 6.5 times the number of DPC equivalent packages as the current LWR fuel cycle. DPC requirement for separated TRISO SNF compared to intact TRISO SNF indicated that a significant reduction in package usage can be accomplished (between 4.2 and 23.8 times) depending on the TRISO loading fractions. This work used the most recent publicly available data from three different advanced reactor vendors. Given the assumptions made and the nature of the data, the results shown here may not reflect the results that each vendor would obtain using their evolving design parameters.

**ENDNOTE**

This is a technical paper that does not take into account contractual limitations or obligations under the Standard Contract for Disposal of Spent Nuclear Fuel and/or High-Level Radioactive Waste (Standard Contract) (10 CFR Part 961). To the extent discussions or recommendations in this paper conflict with the provisions of the Standard Contract, the Standard Contract governs the obligations of the parties, and this paper in no manner supersedes, overrides, or amends the Standard Contract. This paper reflects technical work which could support future decision making by DOE. No inferences should be drawn from this paper regarding future actions by DOE, which are limited both by the terms of the Standard Contract and Congressional appropriations for the Department to fulfill its obligations under the Nuclear Waste Policy Act including licensing and construction of a spent nuclear fuel repository.